\documentclass[amsmath,amssymb,twocolumn,superscriptaddress,
reprint,
 amsmath,amssymb,
 aps,
]{revtex4-2}

\usepackage{graphicx}
\usepackage{dcolumn}
\usepackage{bm}
\usepackage{xcolor} 
\usepackage[normalem]{ulem} 
\usepackage{siunitx}

\newcommand{\psquid}[0]{p-SQUID} 
\newcommand{\fres}[0]{\ensuremath{f_\mathrm{r}}} 
\newcommand{\fqubit}[0]{\ensuremath{f_\mathrm{Q}}} 
\newcommand{\fsq}[0]{\ensuremath{f_\mathrm{pSQ}}} 
\newcommand{\omsq}[0]
{\ensuremath{\omega_\mathrm{pSQ}}} 
\newcommand{\omq}[0]{\ensuremath{\omega_\mathrm{q}}} 
 
\newcommand{\csq}[0]{\ensuremath{C_\mathrm{pSQ}}} 
\newcommand{\gqr}[0]{\ensuremath{g_\mathrm{qr}}} 
\newcommand{\detuning}[0]{\ensuremath{\delta}} 
\newcommand{\phiq}[0]{\ensuremath{\Phi_\mathrm{Q}}} 
\newcommand{\phiext}[0]{\ensuremath{\Phi_\mathrm{ext}}} 
\newcommand{\phisq}[0]{\ensuremath{\Phi_\mathrm{pSQ}}} 
\newcommand{\bpar}[0]{\ensuremath{B_\parallel}} 
\newcommand{\bperp}[0]{\ensuremath{B_\perp}} 
\newcommand{\sq}[0]{\ensuremath{\mathrm{pSQ}}} 
\newcommand{\res}[0]{\ensuremath{\mathrm{r}}} 
\newcommand{\q}[0]{\ensuremath{\mathrm{q}}} 

\begin{document}
\title{Parasitic RF-SQUIDs in superconducting qubits due to wirebonds}

\author{B. Berlitz}
 \affiliation{Physikalisches Institut, Karlsruhe Institute of Technology (KIT), Karlsruhe, Germany}
\author{E. Daum}
 \affiliation{Physikalisches Institut, Karlsruhe Institute of Technology (KIT), Karlsruhe, Germany}
\author{S. Deck}
 \affiliation{Lichttechnisches Institut (LTI), Karlsruhe Institute of Technology (KIT), Karlsruhe, Germany}
 \author{A.V. Ustinov}
 \affiliation{Physikalisches Institut, Karlsruhe Institute of Technology (KIT), Karlsruhe, Germany}
 \affiliation{Institute for Quantum Materials, Karlsruhe Institute of Technology (KIT), Karlsruhe, Germany}
\author{J. Lisenfeld}
\email[corresponding author, eMail: ]{juergen.lisenfeld@kit.edu}

 \affiliation{Physikalisches Institut, Karlsruhe Institute of Technology (KIT), Karlsruhe, Germany}

\date{\today}

\begin{abstract}
Superconducting qubits show great promise to realize practical quantum computers from micro-fabricated integrated circuits.
However, their solid-state architecture bears the burden of parasitic modes in qubit materials and the control circuitry which cause decoherence and interfere with qubits. 
Here, we present evidence that wirebonds, which are used to contact the micro-circuits and to realize chip-to-chip airbridges, may contain parasitic Josephson junctions. In our experiment, such a junction was enclosed in a superconducting loop and so gave rise to the formation an RF-SQUID which interfered with a nearby flux-tunable transmon qubit. 
Periodic signatures observed in magnetic field sweeps revealed a strong AC-dispersive coupling of the parasitic RF-SQUID to both the qubit and its readout resonator, in addition to the DC-inductive coupling between RF-SQUID and qubit.
Our finding sheds light on a previously unknown origin of decoherence due to parasitic Josephson junctions in superconducing circuits. 
\end{abstract}

\maketitle


\section{\label{sec:intro}Introduction}
Superconducting micro-circuits provide a tremendous test-bed to explore quantum coherence in electrically controlled solid-state systems. Different types of custom-tailored quantum two-state-systems (qubits) can be constructed from discrete eigenmodes of resonant circuits which combine the nonlinear inductance of Josephson tunnel junctions with capacitive and inductive elements\cite{reviewsiddiqi2021,reviewkrasnok2024}. 
In recent years, this architecture has become a leading platform in the pursuit to realize large-scale quantum processors~\cite{reviewkjaergaard2020}. Yet, it remains a challenge to keep the macroscopic qubits well isolated from their environment and modes in the controlling circuitry which open channels for decoherence~\cite{Muller:2017,parasiticModeshuang2021}.\\
\begin{figure}[t!]
\centering
\includegraphics[width=\columnwidth]{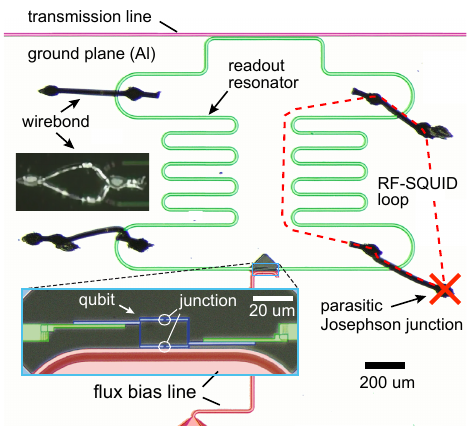}
\caption{False-colored sample micrograph showing the coplanar $\lambda/2$ readout resonator (green) that is coupled at both ends to a mergemon qubit consisting of a DC-SQUID with two Josephson junctions (blue, see inset). Four wirebonds (black) connect the resonator's inner and outer ground planes to avoid slot-line modes. A parasitic Josephson junction within a wirebond (red cross) is enclosed in a superconducting loop (red dashed line) and so gives rise to an RF-SQUID that interferes with qubit and resonator. }
\label{fig:fig1}
\end{figure}
Qubits are typically operated in the microwave regime at resonance frequencies of a few GHz, and read out with weakly coupled and slightly detuned linear resonators that are accessed via on-chip transmission lines~\cite{reviewsiddiqi2021}. Both readout resonators and transmission lines are typically made from coplanar waveguides, where a central signal line is placed between two ground planes.\\
To suppress unwanted slot-line modes, where the currents in the two ground planes oscillate out-of-phase, the state-of-the-art approach is to use micro-fabricated  airbridges~\cite{airbridgesabuwasib2013,airbridgeschen2014,airbridgedunsworth2018} that shorten the ground planes at regular distances. However, in academic research environments and in prototyping experiments, the additional effort of lithographic airbridge fabrication is often spared. Instead, it is common practice to connect the ground plane sections at a few positions by manually placed airbridge wirebonds.\\\vspace{0.1cm}\\
Here, we present evidence that parasitic Josephson junctions (p-JJs) may exist at flawed contacts between aluminum bonding wires and the natively oxidized aluminum film which is part of an on-chip ground plane or bond pad.
This mechanism can give rise to a parasitic RF-SQUID (\psquid) that is formed by the p-JJ together with the inductances of involved wirebonds or lithographic connections which enclose the p-JJ in a superconducting loop.
The quantized magnetic flux in the \psquid\ loop~\cite{RFSQUIDclarke2006} may then couple to neighboring qubits if they are based on SQUIDs, such as flux qubits~\cite{fluxqubitchiorescu2003} and fluxonium qubits\cite{fluxoniummanucharyan2009}, and if they employ DC-SQUIDs for tunability such as transmon qubits~\cite{KochTransmon} and their couplers~\cite{chen2014qubit}.
Moreover, AC-coupling of the \psquid's oscillatory eigenstates may induce dispersive shifts and avoided level crossings in neighboring qubits and resonators.\\

In our experiments, the existence of a \psquid\ was revealed from the characteristic signature it induces in the resonance frequency of a transmon qubit and its readout resonator during sweeps of the applied magnetic field. The data can well be fitted to a model that includes strong inductive and dispersive interactions of the \psquid\ with both the qubit and the readout resonator.\\
The conclusion that a bond wire is involved in the creation of the \psquid\ is supported by measurements with applied in-plane and off-plane magnetic fields which reveal the three-dimensional form of the \psquid\ loop. This is strongly supported by the fact that removing the wirebonds from the sample also removed the p-SQUID signature.
Our finding reveals that parasitic Josephson junctions in the wiring circuitry of superconducting qubits add further decoherence channels for superconducting qubits, and can even completely spoil their operability.

\section{Experiment}
Transmon qubits conventionally consist of one or two submicron-sized Josephson junctions that are shunted by a planar capacitor. In this work, we studied transmon qubits that employ larger-area Josephson junctions which possess sufficient self-capacitance so that no additional shunt capacitor is required. This so-called mergemon design \cite{mergemonzhao2020,mergemonmamin2021} significantly reduces the qubit footprint and promises less interference from charged two-level-system defects (TLS) known to reside at and near the interfaces of qubit electrodes~\cite{Wang2015,Lisenfeld19,Bilmes20,murray2021}. To enable in-situ tuning of the qubit's operation frequency, the junction is replaced by a DC-SQUID that can be flux-biased by a current in a nearby flux bias line. A photograph of the mergemon qubit sample is shown in the inset of Fig.~\ref{fig:fig1}.\\
The qubit's resonance frequency \fqubit\ and state population are read out with a resonator that consists of a coplanar waveguide segment of electrical length $\lambda/2$, whose two open ends are each capacitively coupled to the opposing Mergemon electrodes as shown in Fig.~\ref{fig:fig1}. Since this design choice electrically isolates the inner ground plane of the coplanar resonator, we connected it by four airbridge wirebonds to the surrounding ground plane in order to suppress unwanted slot-line modes. \\

For small qubit-resonator coupling energies $h\gqr$ and sufficiently large detuning $|\detuning| = |(\fqubit - \fres)| \gg \gqr$, the resonator's resonance frequency $\fres$ obtains a dispersive shift\cite{KochTransmon}
\begin{equation}
\fres' = \fres-\frac{\gqr^2}{\detuning}.
\label{eq:dispshift}
\end{equation}
The qubit transition frequency $\fqubit\ $ depends on the flux $\phiq$ threading its DC-SQUID and can be approximated by~\cite{KochTransmon}
\begin{equation}
    h\fqubit(\phiq)=\sqrt{8E_C E_J(\phiq)}-E_C,
    \label{eq:qubitfreq}
\end{equation}
where $E_C = e^2 / 2 C_q$ is the charging energy of the qubit's capacitance $C_q$, $e$ is the electron charge, $E_J = \hbar I_c/2e (\sqrt{\cos(\pi \phiq/\Phi_0)^2+d^2\sin(\pi \phiq/\Phi_0)^2})$ is the Josephson energy assuming two slightly different junctions with a total critical current $I_c$, $\Phi_0=h/2e$ is the flux quantum, $h$ is Planck's constant, and the junction's asymmetry factor $d=\frac{E_{J1}-E_{J2}}{E_{J1}+E_{J2}}$.\\
\begin{figure}
    \centering
    \includegraphics[width=\columnwidth]{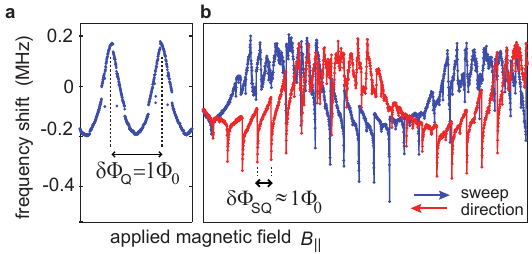}
    \caption{Resonance frequency shift of the readout resonator vs. the applied in-plane magnetic field $B_{\mathrm{||}}$.
    \textbf{a} Sample without chip-to-chip wirebonds, showing the expected modulation as the qubit is tuned according to Eq.~(\ref{eq:qubitfreq}) via the induced flux $\phiq$ in its DC-SQUID. A shift of 0 corresponds to $\fres = 4.5662\,$GHz.
    \textbf{b} With the wirebonds still in place, a saw-tooth pattern and sharp dips in the resonator frequency were additionally observed at regular intervals. 
    Sweeping the applied magnetic field in different directions (blue and red curves) inverts the asymmetry of this signature, and reveals a large hysteresis in the qubit frequency. A shift of 0 corresponds to $\fres = 4.1902\,$GHz.
    }
    \label{fig:fig2}
\end{figure}

Using a network analyzer connected to the on-chip transmission line, the dispersive shift of the readout resonator is determined to estimate the qubit frequency as a function of the applied magnetic fields. In addition to the on-chip flux-bias line, the sample is enclosed in a solenoid to control the magnetic field $\bpar$ in-plane of the qubit circuit, while another coil above the qubit chip applies a magnetic field $\bperp$ perpendicular to the chip's surface.\\
\begin{figure*}[htb]    
    \includegraphics[width=1\textwidth]{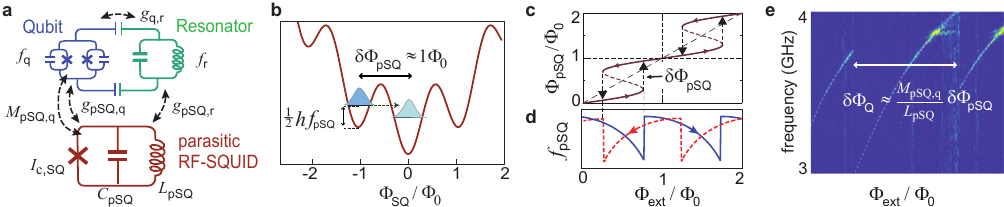}
    \caption{\textbf{a} Circuit schematic including qubit, resonator, and the p-SQUID. Black symbols denote the model parameters. \textbf{b} Potential energy of an RF-SQUID as a function of the flux in its loop $\phisq$. Transitions between wells correspond to a change of $\phisq$ by $\approx 1\Phi_0$.
    \textbf{c} RF-SQUID flux $\phisq$ vs. the applied external flux $\phiext$ for the different sweep directions (arrows). 
    \textbf{d} RF-SQUID resonance frequency $\fsq$ vs. $\phiext$. Blue and red lines indicate different sweep directions.  \textbf{e} The qubit resonance frequency $\fqubit$ (bright pixels), measured vs. the applied magnetic field $\bpar$, shows a sawtooth behaviour due to the discretized flux induced from the \psquid.}
    \label{fig:fig3}
\end{figure*}
Figure~\ref{fig:fig2}\textbf{a} shows the expected periodic oscillations in the resonator's dispersive frequency shift when the qubit is tuned between its lowest and highest (at 4.5662 GHz $\pm$ 200 kHz) resonance frequencies according to Eqs.~(\ref{eq:dispshift}) and (\ref{eq:qubitfreq}) by sweeping the magnetic field $\bpar$. We note that $\bpar$ should nominally be oriented perpendicular to the plane of the qubit's SQUID loop and therefore not couple to $\phiq$. Accordingly, we use these data to characterize the misalignment of the solenoid (see App. \ref{app:geom} for further details).\\

However, the data shown in Fig.~\ref{fig:fig2}\textbf{a} were acquired on the sample after the airbridge wirebonds had been removed. With the wirebonds still in place, the same experiment resulted in the data shown in Fig.~\ref{fig:fig2}\textbf{b}. Most prominently, the resonator frequency displayed a sawtooth-pattern and pronounced asymmetric dips at regular flux intervals in addition to the expected modulation. Moreover, there is a large hysteresis when the direction of the applied field is reversed, which also inverts the asymmetry of the signature.
\section{Model}
In the following, we argue that these observations are explained by the presence of a parasitic RF-SQUID that is formed by a faulty airbridge wirebond that effectively contains a parasitic Josephson junction (p-JJ). We suspect that the p-JJ is contained in one of the four wirebonds that connect inner and outer ground planes of the resonator as shown in Fig.~\ref{fig:fig1}. The dashed red line in Fig.~\ref{fig:fig1} indicates one of the possible superconducting loops that contribute to the \psquid's inductance together with that of other involved wirebonds and on-chip lithographic connections. \\

The schematic of the qubit circuit including the \psquid\ is shown in Fig.~\ref{fig:fig3}\textbf{a}, where $\csq$ is the capacitance of the readout resonator's inner ground plane plus the p-JJ's capacitance.

The potential energy of a large-inductance RF-SQUID is given by the sum of its inductive energy $E_L = \phisq^2 / 2L_\mathrm{SQ}$ and Josephson energy $E_{J\mathrm{,SQ}}(\phisq) \propto \cos(\phisq - \phiext)$ as illustrated in Fig.~\ref{fig:fig3}\textbf{b}.
The height of the potential barrier that localizes the ground state in one of the wells depends on the flux in the \psquid\ loop. Upon increasing the bias flux $\phiext$, the potential barrier will decrease until the state may escape by quantum tunneling or thermal activation to the neighboring deeper well. 
This gives rise to the hysteretic function of $\phisq$ vs. bias flux as illustrated in Fig.~\ref{fig:fig3}\textbf{c}.\\

The nonlinear and quantized flux in the \psquid\ loop $\phisq$ couples via the mutual inductance $M_\mathrm{q,sq}$ to the qubit flux $\phiq$. In magnetic field sweeps, the qubit frequency is thus modulated by a sawtooth-like pattern as it was also observed directly in spectroscopy, see Fig.~\ref{fig:fig3}\textbf{e}. This pattern is reproduced in the dispersive shift of the readout resonator frequency shown in Fig.~\ref{fig:fig2}\textbf{b}. The magnetic coupling between \psquid\ and the qubit also explains the hysteresis and inverted symmetry of the sawtooth pattern when the field sweep direction is reversed. \\
However, the \psquid's oscillatory eigenstates may also be AC-coupled to the qubit and resonator and induce dispersive resonance shifts and avoided level crossings via their respective (transversal) coupling strengths $g_\mathrm{pSQ,q}$ and $g_\mathrm{pSQ,r}$ as described by Eq.~(\ref{eq:dispshift}).\\
The sharp dips observed in $\fres$ (Fig.~\ref{fig:fig2}\textbf{b}) are explained by the dispersive coupling of the \psquid\ to qubit and resonator, which results in a large (quadratic) frequency shift according to Eq.~(\ref{eq:dispshift}) when the detuning becomes small. Figure~\ref{fig:fig3}\textbf{d} shows how the \psquid's plasma frequency $\fsq$ decreases with the tunneling barrier until its state escapes to the next well.
Since only dips but no avoided level crossings nor peaks were observed, the \psquid's resonance frequency must always be above that of the qubit and resonator.
Because the dips remain visible when the qubit is tuned to its lowest frequency, where it is effectively decoupled from the resonator, there must be a strong direct dispersive coupling between the \psquid\ and the resonator, see App.~\ref{app:couplings} for details.
This shows that parasitic RF-SQUIDs can have a strong impact on both qubits and resonators. \\

For a closer test of our model, we numerically simulate the data shown in Fig.~\ref{fig:fig2}\textbf{b}. For each value of bias flux $\phiext$, the small oscillation frequency $\omsq$ in an initial well is found from the \psquid's potential energy \cite{HarlingenSQUID} and
the qubit resonance frequency $\omq$ is calculated from Eq.~(\ref{eq:qubitfreq}), taking into account that the flux in the qubit loop $\phiq$ is a function of the applied field and the flux in the \psquid\ loop $\phisq$. The eigenfrequency of the resonator $\fres$ is then obtained by solving the stationary Schr\"odinger equation for the circuit shown in Fig.~\ref{fig:fig3}\textbf{a} and its Hamiltionian
\begin{equation}
    \frac{\hat{H}(\phi_x)}{h}=\begin{bmatrix}
f_\mathrm{r,0} & g_{\res,\q} & g_{\sq,\res} \\
g_{\q,\res} & f_\mathrm{q,0} & g_{\sq,\q} \\
g_{\sq,\res} & g_{\sq,\q} & f_\mathrm{pSQ,0}\\
\end{bmatrix},
\label{eq:hamiltonian}
\end{equation} which contains the transversal coupling strengths between resonator, qubit, and \psquid\ as off-diagonal elements. The escape from the current potential well is implemented by initializing the \psquid's ground state in the next deeper well once the tunnel rate~\cite{LIKHAREV19811079} exceeds a value of $(20s)^{-1}$ which corresponds to the time required for one measurement of $\fres$.\\

\begin{figure}
\centering
\includegraphics[width=0.9\columnwidth]{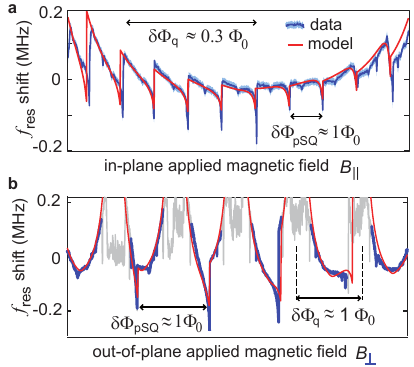}\\
\caption{Dispersive shift of the readout resonator (\textbf{a}) vs. applied in-plane magnetic field $\bpar$, and (\textbf{b}) vs. the applied perpendicular field $\bperp$. The red lines show good agreement with the model's prediction, using the same fitting parameters for both measurements.}
\label{fig:fit}
\end{figure}
Figure~\ref{fig:fit}\textbf{a} shows a striking agreement between the model and the data. To test the hypothesis that a wirebond is involved in the loop of the \psquid, the experiment is repeated by sweeping the out-of-plane component of the applied magnetic field $\bperp$ as shown in Fig.~\ref{fig:fit}\textbf{b}.
A comparison of the flux periodicity of the \psquid\ (given by the distance between dips in $\fres$) with that of the qubit (the distance between the broader maxima in $\fres$) shows that the relative coupling of qubit and p-SQUID to the B-field depends on its direction. This is consistent with the three-dimensional geometry of the wirebond's loops (see App. \ref{app:geom} for details).\\

The model agrees excellently with both experiments for the same set of fit parameters (red lines in Figs.~\ref{fig:fit}\textbf{a} and \textbf{b}). We note that the data obtained near maxima of the qubit frequency is distorted due to the small detuning of qubit and resonator, and has therefore been omitted from the fits.
Since the least-squares fit involved 8 free parameters (see App.~\ref{app:fit}), large uncertainties in the fitting values don't allow us to obtain detailed information on the origin of the parasitic junction that gave rise to the \psquid. 
However, the observed hysteresis clearly reveals the \psquid's multi-well potential associated with~\cite{RFSQUIDclarke2006} $\beta_{L,\sq}=2\pi\, L_\mathrm{\sq}\cdot I_{c,\sq}/\Phi_0>1$. This places the \psquid\ in the parameter regime of flux-biased phase qubits~\cite{phasequbitmartinis2003,phasequbitmartinis2009,phasequbitlisenfeld2008}.
We estimate the possible ranges of \psquid\ parameters by fixing $\beta_{L,\sq}$ to a value of 11.3 which is compatible with the observed hysteresis. The restricted fit (see Appendix~\ref{app:wireinductance}) achieves good agreement with the data when $L_\sq$ ranges between $0.1$ and 1.7 nH, corresponding to critical currents of $I_{c,\sq}$ between $80\mu A$ and $20\mu A$ and capacitances  $C_{\sq}$ of 30pF to 5 pF. While the best-fitting inductance of $L_{\sq}\approx474$ pH is well in accordance with the wirebonds' inductances, the capacitance and critical current of the parasitic junction can not be predicted without exact knowledge of its geometry since it depends strongly on the dielectric material (such as photoresist residual) and its thickness.
\\
Aiming to identify the flawed bond after the sample was warmed up, the bond wire bridges were touched and heavily bent using the micro-manipulator of a bonding station. However, all bonds appeared stable and did not disconnect easily. We note that it has been proposed to utilize the discreteness of RF-SQUID flux states to provide a stable flux-bias source for qubits~\cite{RFSQUIDcastellano2006, rfSQUIDtuning}.\\

\section{Discussion}
The observed anomaly of resonator and qubit resonance frequencies in applied magnetic fields is fully explained by their interactions with a parasitic RF-SQUID. The notion that a faulty wirebond contains a parasitic tunnel junction is supported by in- and out-of-plane magnetic field sweeps which confirm the three-dimensional form of the \psquid's loop. A further confirmation is the fact that the anomaly disappeared once the wirebonds were removed. Other possible origins of flux irregularities, such as the motion of trapped vortices~\cite{VortexHopping}, would not result in the observed strictly periodic signatures and observed dispersive resonance frequency shifts.\\

While it has been known that stray Josephson junctions between thin-film qubit electrodes are a source of qubit decoherence~\cite{dunsworth2017,Lisenfeld19}, our finding sheds light on related issues in the remaining circuitry and packaging. In our experiment, two out of four qubits on the chip were inoperable due to their strong interference with the parasitic RF-SQUID formed by a faulty chip-to-chip wirebond.\\
The signature of a parasitic SQUID was so far only observed in a single experiment. To determine how common this problem actually is will require a systematic study on many samples, in which a large range of applied magnetic fields is explored to search for anomalies such as avoided level crossings in resonator and qubit spectroscopy.
It can be seen as a coincidence that the parameters of the \psquid\ placed it into a regime where its interactions with qubit and resonator could be clearly observed. In more typical cases, the \psquid\ could be further detuned and induce only slight frequency shifts in a circuit that may remain unnoticed. Similarly, \psquid s may be overlooked when only on-chip qubit flux bias lines are used that only slightly tune the \psquid's resonance frequency, while their parasitic modes would still contribute to decoherence and flux noise.\\

Related issues may arise for instance from detrimental modes in flux bias lines that were terminated by a junction in a flawed wirebond. Problems with parasitic junctions due to flawed airbridges, bond wires, or bump-pads in flip-chip architectures\cite{flipchiprosenberg20173}, may rise in importance as quantum processors increase in size and complexity. The observed strong DC-magnetic and AC-electric interactions of parasitic RFSQUIDs can furthermore spoil the functionality of various other superconducting devices~\cite{reviewbraginski2019} such as for instance parametric amplifiers\cite{twpamacklin2015near,twpareviewaumentado2020}, RSFQ-logic~\cite{rsfqbunykreview}, MKID photon detectors\cite{MKIDreviewmazin2009}, and passive resonator-based microwave filters~\cite{filtersoates2022}.\\

\begin{acknowledgments}
 We acknowledge the contributions of Alexander Bilmes to the experimental setup and qubit fabrication.  We thank Hannes Rotzinger for fruitful discussions and vital help in the laboratory and clean-room. BB acknowledges help of Konstantin Händel with sample fabrication. This work was funded by Google, which we gratefully acknowledge.
 \end{acknowledgments}

\appendix

\section{\label{app:fit}Fit to the model}
To compare the data shown in Figs.~\ref{fig:fit} to the system of interacting qubit, resonator, and \psquid\ , we simulate magnetic field sweeps and obtain the eigenfrequency of the resonator by numerical diagonalization of the Hamiltonian Eq.~(\ref{eq:hamiltonian}) as described in the main text. Best-fitting model parameters are then obtained by a least-square fit routine. These are listed in Table~\ref{table:fitparms} and were used to plot the red lines in Fig.~\ref{fig:fit}.\\
However, due to the large number and not fully independent model parameters, the confidence interval of most fitting parameters greatly exceeds their values so that we cannot obtain precise information on the \psquid's parameters and geometry. This could be improved with the help of finite element simulations of the \psquid's inductance and capacitance. \\

\begin{table} [htb]
    \centering
    \begin{tabular}{llll}
    parameter & value & description\\ 
    \hline 
    $f_\mathrm{res,0}$ & 4.19 GHz & uncoupled resonator frequency  \\ 
    $g_{\sq,\res}$ & 9.9 MHz & coupling strength pSQUID-resonator \\ 
    $g_{\sq,\q}$ & 70.3 MHz & coupling strength pSQUID-qubit \\ 
    $\phi_{slope}^Q$ & 0.177 $\Phi_0$ &\phiq induced between well jumps\\ 
    $\phi_{qjump}^Q$ & 0.116 $\Phi_0$&\phiq jump at well jump\\ 
    $L_\sq$  & 474 pH & \psquid\ inductance  \\ 
    $C_\sq$  & 10.3 pF & \psquid\ capacitance\\ 
    $I_{c,\sq}$  & 7.85 $\mu$A & \psquid\ critical current \\ 
    \end{tabular}
 \caption{Model parameters and their values that best fit the data as shown in Fig.~\ref{fig:fit}. We note that due to large uncertainties in the fitting values, the fit only serves for a qualitative comparison with the model.}
    \label{table:fitparms}
\end{table}
The parameters of the qubit and its coupling to the resonator (see Table~\ref{table:measparms}) were obtained from a fit to spectroscopy data on the sample after the wirebonds had been removed. We note that the readout resonator's frequency was 4.5662 GHz when the wirebonds were removed, and 4.1902 GHz with installed airbridge wirebonds as shown in Fig.~\ref{fig:fig1}.\\ 

\begin{table}[htb]
    \centering
    \begin{tabular}{llll}
    parameter & value & description\\ 
    \hline 
    $E_J$ &12.3 GHz  &qubit Josephson energy  &   \\ 
    {$f_\mathrm{q,max}$} & 4.125 GHz & maximum qubit frequency at zero flux\\
    $E_C$ &0.19 GHz  &qubit charge energy  &  \\ 
    $d$  & 0.12 &qubit JJ asymmetry factor  &  \\ 
    $g_{\q,\res}$ &0.014 GHz  &  coupling strength qubit-resonator  \\    
    \hline 
    \end{tabular}
    \caption{Qubit parameters, obtained in spectroscopy after removal of wirebonds.}
    \label{table:measparms}
\end{table}

To estimate the range of pSQUID parameters which result in reasonable agreement with the data, we fixed the parameter $\beta_L= 2\pi L_\mathrm{pSQ} I_\mathrm{C,pSQ} / \Phi_0$ to the value of 11.3 to ensure that the p-SQUID hysteresis matches the data shown in Fig.~\ref{fig:fig2}.
A value of the pSQUID inductance $L_{\mathrm{pSQ}}$ was then chosen, and the pJJ's critical current $I_\mathrm{C,pSQ}$ was calculated accordingly. The other parameters listed in Table \ref{table:fitparms} were kept as free fitting parameters.\\
Figure \ref{figure:reduced_fit}b shows the deviation from the data in the form of the root mean square error of the fit. The fit quality remains rather constant for a wide range of $L_{\mathrm{pSQ}}$ between 0.1-1.7 nH. For higher values, the fit quickly becomes worse as it becomes impossible to fulfill the three boundary conditions imposed by the p-SQUID-hysteresis, the fixed $L_{\mathrm{pSQ}}$, and the requirement that the pSQUID's resonance frequency must always exceed that of the readout resonator. \\
The uncertainties in the p-SQUID capacitance $C_\sq$, and the parameters controlling the p-SQUID-qubit flux-coupling $\phi_{slope}^Q$, $\phi_{qjump}^Q$, range between 5-12$\%$. This is a much more narrow uncertainty range compared to a fit where all parameters are left free. The uncertainty in the dispersive coupling strengths $g_{\sq,\res}$ and $g_{\sq,\q}$ remains large (20 to 90\%) because both contribute to the formation of the sharp dips in the resonator frequency.\\

\section{\label{app:couplings}\psquid\ coupling to qubit and resonator}
\begin{figure}
    \centering
    \includegraphics[width=1\columnwidth]{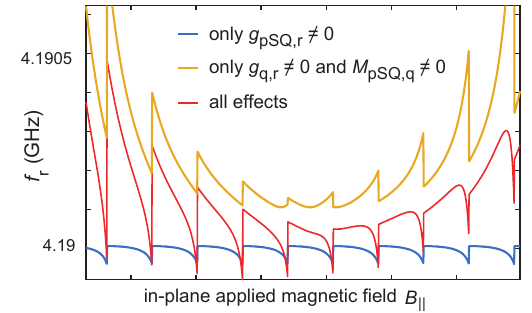}
    \caption{Resonator resonance frequency $\fres$ vs. $\bpar$ obtained from the Hamiltonian Eq.~\ref{eq:hamiltonian} when different coupling parameters are set to zero. Without a direct transversal coupling between \psquid\ and the resonator (yellow line), the sharp dips near the minimum of $\fres$ are absent. The blue line shows the dips induced by this direct interaction.
    }
    \label{fig:app_comparison}
\end{figure}

To illustrate our conclusion that the \psquid\ directly couples to the qubit's readout resonator, we set different parameters in the Hamiltonian Eq.~(\ref{eq:hamiltonian}) to zero and compare the model's prediction shown in Fig.~\ref{fig:app_comparison}. This shows that the sharp dips in $\fres$ are caused by the dispersive shift on $\fres$ due to nonzero \psquid-resonator transversal coupling $g_\mathrm{pSQ,r}$.

\section{Estimating the pSQUID parameters}\label{app:wireinductance}
To estimate the range of realistic values for the inductance of the pSQUID, we consider the geometry of the wirebonds shown in Fig.~\ref{fig:fig1}. From optical microscopy as shown in Fig.~\ref{figure:wireinductance}a, the lengths of the four wirebonds is measured as 381\textmu m, 466\textmu m, 466\textmu m, and  546\textmu m, respectively. Using the textbook formula~\cite{rosa1908self} 
\begin{equation}
    L=\frac{\mu_0}{2\pi}l\left(\log\left(\frac{2l}{r}\right)-1\right),
\end{equation}
where $r$ is the wire radius and $l$ is its length, we estimate inductances of 237pH to 379pH for our aluminum bond wires which have a diameter of 25\textmu m. In Fig.~\ref{figure:wireinductance}b, the inductance per length is plotted also for other wire diameters, and compared to a common rule of thumb which estimates that a 1mm-long bond wire has a typical inductance of 1 nH.\\

\begin{figure}[h]
\centering
\includegraphics[width=\columnwidth]{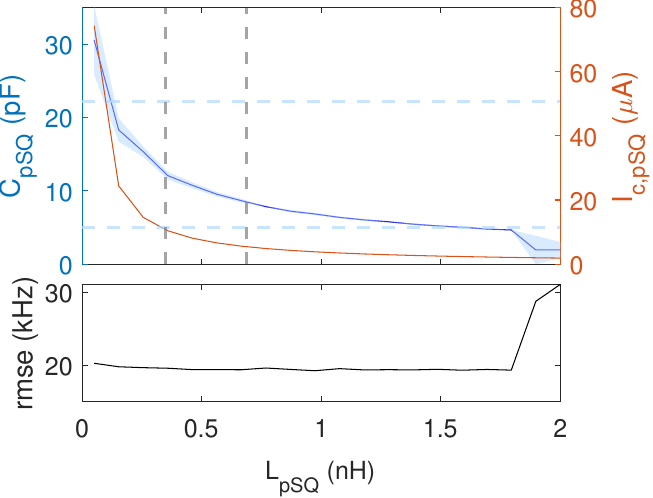}
\caption{Top panel: Fitting values of $C_\mathrm{pSQ}$ and $I_{c,\sq}$ as a function of $L_{\mathrm{pSQ}}$
for fixed $\beta_L \equiv2\pi L I_C/\Phi_0 = 11.3$ (as determined by a measurement of the \psquid's hysteresis).
The bottom panel shows the root mean squared error of the fit, which remains similarly small in a wide range of $L_{\mathrm{pSQ}}$. The vertical dashed lines in the top panel indicate the realistic range of bond wire inductances as described in the text.}
\label{figure:reduced_fit}
\end{figure}

To obtain a lower limit for the range of reasonable pSQUID inductance, we assume that the parasitic junction is contained in the shortest bond wire, and that the pSQUID loop is closed by the parallel connection of the remaining three wirebonds. 
The upper limit is estimated by assuming that only the two longest bond wires are involved in the loop formation. This results in an estimated range of $347 \mathrm{pH} \lesssim L_\mathrm{pSQ} \lesssim 688$ pH, which is well in accordance with the best fitting parameter $L_\mathrm{pSQ}$ of 474 pH. The limits are indicated by vertical dashed lines in Fig. \ref{figure:reduced_fit}.\\

For the pSQUID capacitance, an upper limit can be estimated from the average area of a wirebond foot that is measured to about $A_\mathrm{foot}\approx (62\,\mu m)^2 =3840\, \mu\mathrm{m}^2$. Taking $d=2\,$nm as a minimal thickness of the capacitor dielectric (so that the whole bond wire foot would form a tunnel junction) and a relative permittivity $\varepsilon_r \approx 10$ for AlOx, we obtain $C_\mathrm{pSQ} = \epsilon_0 \epsilon_r A_\mathrm{foot}/d =$170 pF which is much larger than the fitted $C_\mathrm{pSQ} \approx 10$pF. Realistically, the whole bond foot area must contribute to $C_\sq$. The discrepancy between this large area and relatively small (fitting) capacitance can be explained by a larger thickness and lower permittivity of the dielectric under the faulty bonding wire foot. For example, assuming that the bond foot area is contaminated by a 10 nm-thick layer of residual PMMA photoresist ($\varepsilon_r\approx3$) results in a capacitance of 10 pF that corresponds to the best fitting value.\\

\begin{figure}
\centering
\includegraphics[width=6.5cm]{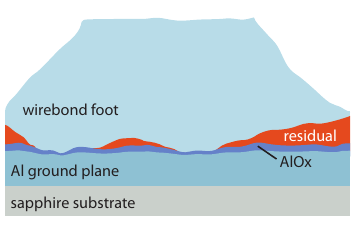}
\caption{Sketch of the cross-section through the foot of a wirebond (not to scale), illustrating thickness variations of the native AlOx on the ground plane and a layer of contamination such as photoresist residual. The active junction area where Cooper pair tunneling occurs may be limited to small regions of least dielectric thickness in addition to pinholes.}
\label{fig:app_footjj}
\end{figure}

\begin{figure}
\centering
\includegraphics[width=\columnwidth]{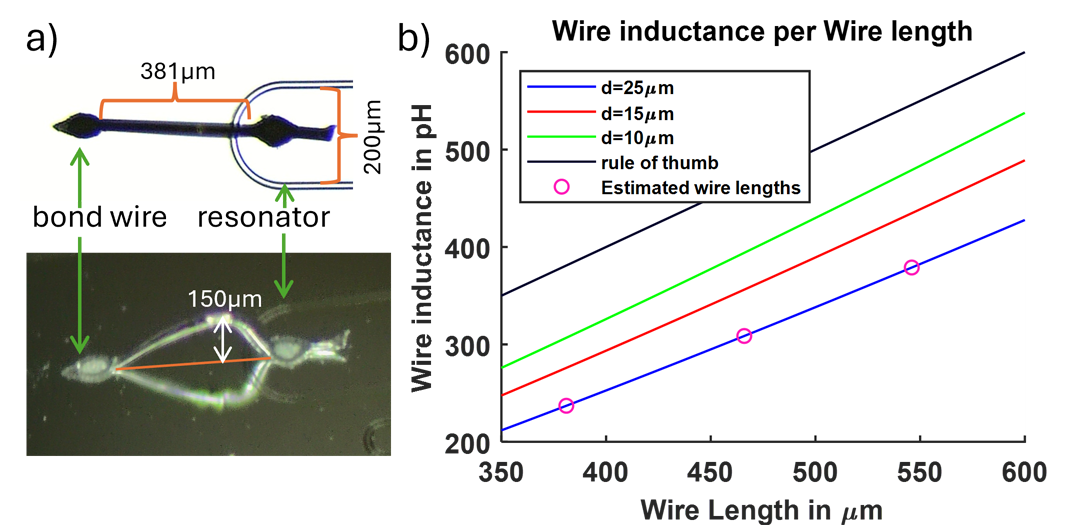}
\caption{a) Optical micrographs of bonding wires. By comparing micrographs from different angles, the lengths of the bonding wires were measured to be 466, 381, 546 and 466 $\mu m$. b) The bond wire inductance per length of wire calculated by the textbook formula for different wire diameters (blue, red and green line). }
\label{figure:wireinductance}
\end{figure}

To reconcile the large bond foot area with the small capacitance (requiring a thick dielectric) and formation of a tunnel junction (requiring a tunnel barrier thickness of $\lesssim 3$nm), one can assume that the thickness of the dielectric under the wirebond foot is non-uniform, possibly due to photoresist residuals.  Figure \ref{fig:app_footjj} illustrates this situation in a cross-section through the wirebond foot. The actual contamination of our sample surface is unknown. In ultrasonic wirebonding, the amount of intermetallic formed at the bonded interface depends strongly on the bonding parameters such as ultrasound frequency, amplitude, and duration~\cite{Prasad2004}, so that inadequate settings could have played a role in the junction formation.\\

Since the Cooper-pair tunneling rate depends exponentially on the tunnel barrier thickness, the critical current of the parasitic junction can not be predicted without exact knowledge of the junction geometry. The junction's active tunneling area might be given by a small region of least surface contamination and be could be largely affected by pinholes.
In the restricted fit, compatible fit values for the critical current range between 2 and 80 $\mu$A as shown in Fig.~\ref{figure:reduced_fit}.\\

\section{Field alignment and p-SQUID geometry}
\label{app:geom}
\begin{figure}
\centering
\includegraphics[width=0.7\columnwidth]{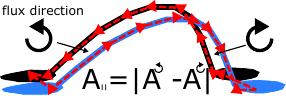}
\caption{Illustration of adjacent wirebond loops from an in-plane perspective. Their effective loop size A is the difference of the area enclosed by loop segments of opposite current directions.}
\label{figure:SquidLoop}
\end{figure}

To estimate the misalignment of the in-plane magnetic field with respect to the chip's ground plane, we measure the periodicity of the qubit resonance frequency as a function of $\bpar$. Using the area of the qubit's DC-SQUID loop $A=276\,\mu m^2$, we calculate the misalignment angle $\alpha$ from the relation
\begin{equation}
    \alpha=\sin^{-1}\left(\frac{\Phi_0}{AB_{||}}\right)=0.0725^{\circ},
\end{equation}
where the field $\bpar$ generated by the solenoid is estimated from COMSOL simulations.\\
The \psquid's effective loop area $A_{pSQ,\parallel}$ can accordingly be estimated from the periodicity of the dips in the resonator frequency (Fig.~\ref{fig:fit}) as a function of the estimated $\bpar$-field strength. This results in
\begin{equation}
    A_{pSQ,\parallel}=\frac{\Phi_0}{B_{||}^{0}}=2068\,\mu m^2.
\end{equation}
   The effective \psquid\ area is rather small due to gradiometer effects in the four parallel wirebond connections as shown in Fig.~\ref{fig:fig1} and the illustration in Fig.~\ref{figure:SquidLoop}.\\

\section{Sample fabrication}\label{app:samplefabrication}
The samples were fabricated on a C-plane oriented and \SI{500}{\micro \metre} thick 3-inch sapphire wafer. After cleaning in Piranha solution and an oxygen plasma, it is installed in a \textit{PLASYS MEB 550 S} eBeam evaporation chamber, heated to $\SI{200}{\celsius}$ for two hours to remove any excess moisture, and subsequently evaporated with $\SI{100}{\nano \metre}$ aluminum at a rate of $\SI{1}{\nano \metre \per \second}$, a polar angle of $\ang{0}$, and a base pressure of $5 \times 10^{-8}\si{\milli \bar}$ to form the ground plane. The aluminum layer is then passivated by means of a static oxidation at $\SI{30}{\milli \bar}$ for $\SI{10}{\minute}$ and exposed to the clean-room atmosphere.\\
This aluminum ground plane is then structured using optical lithography. For this purpose, \textit{S1805} photoresist is applied and patterned using a mask-aligner and \textit{AZ Developer}, followed by Al etching in an inductively coupled plasma. Afterwards, the resist is removed and the wafer cleaned with DMSO, 2-propanol and an oxygen plasma.
\\
The Josephson junctions and leads were fabricated using a modified version of the \textit{in-situ} bandaged Niemeyer-Dolan technique that avoids the formation of stray junctions \cite{Bilmes21}. For this, a bi-layer of $\SI{250}{\nano \metre}$ \textit{A4} (PMMA) on top of $\SI{900}{\nano \metre}$ \textit{EL-13} (MMA) was applied to the sample and patterned using an electron beam writing system and developed with a mixture of 2-propanol and bidistilled water. Resist residuals in the trenches were removed using an oxygen plasma. Afterwards, the Josephson junctions and leads were deposited in the \textit{PLASYS MEB 550 S} at a base pressure of $5 \times 10^{-8}\si{\milli \bar}$.\\
Deviating from conventional fabrication, the here-used samples employed thicker-than-usual tunnel barriers. These were realized by first oxidizing the deposited junction bottom electrodes at a static pressure of $\SI{130}{\milli \bar}$ for $\SI{20}{\minute}$. To further thicken the barrier, an additional aluminium layer of about $\SI{0.1}{\nano \metre}$ thickness was deposited and thoroughly oxidized, followed by deposition of the junction's top electrode. After protective oxidation at $\SI{30}{\milli \bar}$ for $\SI{10}{\minute}$, a lift-off was performed using DMSO and the sample is cleaned in 2-propanol and an ultrasonic bath. The resulting tunnel junctions had critical current densities of about 2 A/cm$^2$. A thorough discussion of junction fabrication and qubit coherence will be the focus of an upcoming work~\cite{Daum25}.

\section{Experimental setup}
The sample was measured in a dry dilution refrigerator at a base temperature of about 25 mK. The readout resonator frequency was measured with a vector network analyzer via a transmission line that had total attenuation of 80 dB distributed over several temperature stages, and was equipped with band-pass and infrared filters. After interaction with the resonator, the signal was isolated by means of two circulators and additional band-pass and infrared filters from a HEMT amplifier at the 4K stage, and further amplified by 66 dB at room temperature.


\begin{thebibliography}{39}%
	\makeatletter
	\providecommand \@ifxundefined [1]{%
		\@ifx{#1\undefined}
	}%
	\providecommand \@ifnum [1]{%
		\ifnum #1\expandafter \@firstoftwo
		\else \expandafter \@secondoftwo
		\fi
	}%
	\providecommand \@ifx [1]{%
		\ifx #1\expandafter \@firstoftwo
		\else \expandafter \@secondoftwo
		\fi
	}%
	\providecommand \natexlab [1]{#1}%
	\providecommand \enquote  [1]{``#1''}%
	\providecommand \bibnamefont  [1]{#1}%
	\providecommand \bibfnamefont [1]{#1}%
	\providecommand \citenamefont [1]{#1}%
	\providecommand \href@noop [0]{\@secondoftwo}%
	\providecommand \href [0]{\begingroup \@sanitize@url \@href}%
	\providecommand \@href[1]{\@@startlink{#1}\@@href}%
	\providecommand \@@href[1]{\endgroup#1\@@endlink}%
	\providecommand \@sanitize@url [0]{\catcode `\\12\catcode `\$12\catcode
		`\&12\catcode `\#12\catcode `\^12\catcode `\_12\catcode `\%12\relax}%
	\providecommand \@@startlink[1]{}%
	\providecommand \@@endlink[0]{}%
	\providecommand \url  [0]{\begingroup\@sanitize@url \@url }%
	\providecommand \@url [1]{\endgroup\@href {#1}{\urlprefix }}%
	\providecommand \urlprefix  [0]{URL }%
	\providecommand \Eprint [0]{\href }%
	\providecommand \doibase [0]{https://doi.org/}%
	\providecommand \selectlanguage [0]{\@gobble}%
	\providecommand \bibinfo  [0]{\@secondoftwo}%
	\providecommand \bibfield  [0]{\@secondoftwo}%
	\providecommand \translation [1]{[#1]}%
	\providecommand \BibitemOpen [0]{}%
	\providecommand \bibitemStop [0]{}%
	\providecommand \bibitemNoStop [0]{.\EOS\space}%
	\providecommand \EOS [0]{\spacefactor3000\relax}%
	\providecommand \BibitemShut  [1]{\csname bibitem#1\endcsname}%
	\let\auto@bib@innerbib\@empty
	\bibitem [{\citenamefont {Siddiqi}(2021)}]{reviewsiddiqi2021}%
	\BibitemOpen
	\bibfield  {author} {\bibinfo {author} {\bibfnamefont {I.}~\bibnamefont
			{Siddiqi}},\ }\bibfield  {title} {\bibinfo {title} {Engineering
			high-coherence superconducting qubits},\ }\href@noop {} {\bibfield  {journal}
		{\bibinfo  {journal} {Nature Reviews Materials}\ }\textbf {\bibinfo {volume}
			{6}},\ \bibinfo {pages} {875} (\bibinfo {year} {2021})}\BibitemShut {NoStop}%
	\bibitem [{\citenamefont {Krasnok}\ \emph {et~al.}(2024)\citenamefont
		{Krasnok}, \citenamefont {Dhakal}, \citenamefont {Fedorov}, \citenamefont
		{Frigola}, \citenamefont {Kelly},\ and\ \citenamefont
		{Kutsaev}}]{reviewkrasnok2024}%
	\BibitemOpen
	\bibfield  {author} {\bibinfo {author} {\bibfnamefont {A.}~\bibnamefont
			{Krasnok}}, \bibinfo {author} {\bibfnamefont {P.}~\bibnamefont {Dhakal}},
		\bibinfo {author} {\bibfnamefont {A.}~\bibnamefont {Fedorov}}, \bibinfo
		{author} {\bibfnamefont {P.}~\bibnamefont {Frigola}}, \bibinfo {author}
		{\bibfnamefont {M.}~\bibnamefont {Kelly}},\ and\ \bibinfo {author}
		{\bibfnamefont {S.}~\bibnamefont {Kutsaev}},\ }\bibfield  {title} {\bibinfo
		{title} {Superconducting microwave cavities and qubits for quantum
			information systems},\ }\href@noop {} {\bibfield  {journal} {\bibinfo
			{journal} {Applied Physics Reviews}\ }\textbf {\bibinfo {volume} {11}}
		(\bibinfo {year} {2024})}\BibitemShut {NoStop}%
	\bibitem [{\citenamefont {Kjaergaard}\ \emph {et~al.}(2020)\citenamefont
		{Kjaergaard}, \citenamefont {Schwartz}, \citenamefont {Braum{\"u}ller},
		\citenamefont {Krantz}, \citenamefont {Wang}, \citenamefont {Gustavsson},\
		and\ \citenamefont {Oliver}}]{reviewkjaergaard2020}%
	\BibitemOpen
	\bibfield  {author} {\bibinfo {author} {\bibfnamefont {M.}~\bibnamefont
			{Kjaergaard}}, \bibinfo {author} {\bibfnamefont {M.~E.}\ \bibnamefont
			{Schwartz}}, \bibinfo {author} {\bibfnamefont {J.}~\bibnamefont
			{Braum{\"u}ller}}, \bibinfo {author} {\bibfnamefont {P.}~\bibnamefont
			{Krantz}}, \bibinfo {author} {\bibfnamefont {J.~I.-J.}\ \bibnamefont {Wang}},
		\bibinfo {author} {\bibfnamefont {S.}~\bibnamefont {Gustavsson}},\ and\
		\bibinfo {author} {\bibfnamefont {W.~D.}\ \bibnamefont {Oliver}},\ }\bibfield
	{title} {\bibinfo {title} {Superconducting qubits: Current state of play},\
	}\href@noop {} {\bibfield  {journal} {\bibinfo  {journal} {Annual Review of
				Condensed Matter Physics}\ }\textbf {\bibinfo {volume} {11}},\ \bibinfo
		{pages} {369} (\bibinfo {year} {2020})}\BibitemShut {NoStop}%
	\bibitem [{\citenamefont {Müller}\ \emph {et~al.}(2019)\citenamefont
		{Müller}, \citenamefont {Cole},\ and\ \citenamefont
		{Lisenfeld}}]{Muller:2017}%
	\BibitemOpen
	\bibfield  {author} {\bibinfo {author} {\bibfnamefont {C.}~\bibnamefont
			{Müller}}, \bibinfo {author} {\bibfnamefont {J.~H.}\ \bibnamefont {Cole}},\
		and\ \bibinfo {author} {\bibfnamefont {J.}~\bibnamefont {Lisenfeld}},\
	}\bibfield  {title} {\bibinfo {title} {Towards understanding
			two-level-systems in amorphous solids: insights from quantum circuits},\
	}\href {https://doi.org/10.1088/1361-6633/ab3a7e} {\bibfield  {journal}
		{\bibinfo  {journal} {Reports on Progress in Physics}\ }\textbf {\bibinfo
			{volume} {82}},\ \bibinfo {pages} {124501} (\bibinfo {year}
		{2019})}\BibitemShut {NoStop}%
	\bibitem [{\citenamefont {Huang}\ \emph {et~al.}(2021)\citenamefont {Huang},
		\citenamefont {Lienhard}, \citenamefont {Calusine}, \citenamefont
		{Veps{\"a}l{\"a}inen}, \citenamefont {Braum{\"u}ller}, \citenamefont {Kim},
		\citenamefont {Melville}, \citenamefont {Niedzielski}, \citenamefont {Yoder},
		\citenamefont {Kannan} \emph {et~al.}}]{parasiticModeshuang2021}%
	\BibitemOpen
	\bibfield  {author} {\bibinfo {author} {\bibfnamefont {S.}~\bibnamefont
			{Huang}}, \bibinfo {author} {\bibfnamefont {B.}~\bibnamefont {Lienhard}},
		\bibinfo {author} {\bibfnamefont {G.}~\bibnamefont {Calusine}}, \bibinfo
		{author} {\bibfnamefont {A.}~\bibnamefont {Veps{\"a}l{\"a}inen}}, \bibinfo
		{author} {\bibfnamefont {J.}~\bibnamefont {Braum{\"u}ller}}, \bibinfo
		{author} {\bibfnamefont {D.~K.}\ \bibnamefont {Kim}}, \bibinfo {author}
		{\bibfnamefont {A.~J.}\ \bibnamefont {Melville}}, \bibinfo {author}
		{\bibfnamefont {B.~M.}\ \bibnamefont {Niedzielski}}, \bibinfo {author}
		{\bibfnamefont {J.~L.}\ \bibnamefont {Yoder}}, \bibinfo {author}
		{\bibfnamefont {B.}~\bibnamefont {Kannan}}, \emph {et~al.},\ }\bibfield
	{title} {\bibinfo {title} {Microwave package design for superconducting
			quantum processors},\ }\href@noop {} {\bibfield  {journal} {\bibinfo
			{journal} {PRX Quantum}\ }\textbf {\bibinfo {volume} {2}},\ \bibinfo {pages}
		{020306} (\bibinfo {year} {2021})}\BibitemShut {NoStop}%
	\bibitem [{\citenamefont {Abuwasib}\ \emph {et~al.}(2013)\citenamefont
		{Abuwasib}, \citenamefont {Krantz},\ and\ \citenamefont
		{Delsing}}]{airbridgesabuwasib2013}%
	\BibitemOpen
	\bibfield  {author} {\bibinfo {author} {\bibfnamefont {M.}~\bibnamefont
			{Abuwasib}}, \bibinfo {author} {\bibfnamefont {P.}~\bibnamefont {Krantz}},\
		and\ \bibinfo {author} {\bibfnamefont {P.}~\bibnamefont {Delsing}},\
	}\bibfield  {title} {\bibinfo {title} {Fabrication of large dimension
			aluminum air-bridges for superconducting quantum circuits},\ }\href@noop {}
	{\bibfield  {journal} {\bibinfo  {journal} {Journal of Vacuum Science \&
				Technology B}\ }\textbf {\bibinfo {volume} {31}} (\bibinfo {year}
		{2013})}\BibitemShut {NoStop}%
	\bibitem [{\citenamefont {Chen}\ \emph
		{et~al.}(2014{\natexlab{a}})\citenamefont {Chen}, \citenamefont {Megrant},
		\citenamefont {Kelly}, \citenamefont {Barends}, \citenamefont {Bochmann},
		\citenamefont {Chen}, \citenamefont {Chiaro}, \citenamefont {Dunsworth},
		\citenamefont {Jeffrey}, \citenamefont {Mutus} \emph
		{et~al.}}]{airbridgeschen2014}%
	\BibitemOpen
	\bibfield  {author} {\bibinfo {author} {\bibfnamefont {Z.}~\bibnamefont
			{Chen}}, \bibinfo {author} {\bibfnamefont {A.}~\bibnamefont {Megrant}},
		\bibinfo {author} {\bibfnamefont {J.}~\bibnamefont {Kelly}}, \bibinfo
		{author} {\bibfnamefont {R.}~\bibnamefont {Barends}}, \bibinfo {author}
		{\bibfnamefont {J.}~\bibnamefont {Bochmann}}, \bibinfo {author}
		{\bibfnamefont {Y.}~\bibnamefont {Chen}}, \bibinfo {author} {\bibfnamefont
			{B.}~\bibnamefont {Chiaro}}, \bibinfo {author} {\bibfnamefont
			{A.}~\bibnamefont {Dunsworth}}, \bibinfo {author} {\bibfnamefont
			{E.}~\bibnamefont {Jeffrey}}, \bibinfo {author} {\bibfnamefont
			{J.}~\bibnamefont {Mutus}}, \emph {et~al.},\ }\bibfield  {title} {\bibinfo
		{title} {Fabrication and characterization of aluminum airbridges for
			superconducting microwave circuits},\ }\href@noop {} {\bibfield  {journal}
		{\bibinfo  {journal} {Applied Physics Letters}\ }\textbf {\bibinfo {volume}
			{104}} (\bibinfo {year} {2014}{\natexlab{a}})}\BibitemShut {NoStop}%
	\bibitem [{\citenamefont {Dunsworth}\ \emph {et~al.}(2018)\citenamefont
		{Dunsworth}, \citenamefont {Barends}, \citenamefont {Chen}, \citenamefont
		{Chen}, \citenamefont {Chiaro}, \citenamefont {Fowler}, \citenamefont
		{Foxen}, \citenamefont {Jeffrey}, \citenamefont {Kelly}, \citenamefont
		{Klimov} \emph {et~al.}}]{airbridgedunsworth2018}%
	\BibitemOpen
	\bibfield  {author} {\bibinfo {author} {\bibfnamefont {A.}~\bibnamefont
			{Dunsworth}}, \bibinfo {author} {\bibfnamefont {R.}~\bibnamefont {Barends}},
		\bibinfo {author} {\bibfnamefont {Y.}~\bibnamefont {Chen}}, \bibinfo {author}
		{\bibfnamefont {Z.}~\bibnamefont {Chen}}, \bibinfo {author} {\bibfnamefont
			{B.}~\bibnamefont {Chiaro}}, \bibinfo {author} {\bibfnamefont
			{A.}~\bibnamefont {Fowler}}, \bibinfo {author} {\bibfnamefont
			{B.}~\bibnamefont {Foxen}}, \bibinfo {author} {\bibfnamefont
			{E.}~\bibnamefont {Jeffrey}}, \bibinfo {author} {\bibfnamefont
			{J.}~\bibnamefont {Kelly}}, \bibinfo {author} {\bibfnamefont
			{P.}~\bibnamefont {Klimov}}, \emph {et~al.},\ }\bibfield  {title} {\bibinfo
		{title} {A method for building low loss multi-layer wiring for
			superconducting microwave devices},\ }\href@noop {} {\bibfield  {journal}
		{\bibinfo  {journal} {Applied Physics Letters}\ }\textbf {\bibinfo {volume}
			{112}} (\bibinfo {year} {2018})}\BibitemShut {NoStop}%
	\bibitem [{\citenamefont {Clarke}\ and\ \citenamefont
		{Braginski}(2006)}]{RFSQUIDclarke2006}%
	\BibitemOpen
	\bibfield  {author} {\bibinfo {author} {\bibfnamefont {J.}~\bibnamefont
			{Clarke}}\ and\ \bibinfo {author} {\bibfnamefont {A.~I.}\ \bibnamefont
			{Braginski}},\ }\href@noop {} {\emph {\bibinfo {title} {The SQUID handbook.
				Vol. 2. Applications of SQUIDs and SQUID systems}}}\ (\bibinfo  {publisher}
	{Wiley-VCH},\ \bibinfo {year} {2006})\BibitemShut {NoStop}%
	\bibitem [{\citenamefont {Chiorescu}\ \emph {et~al.}(2003)\citenamefont
		{Chiorescu}, \citenamefont {Nakamura}, \citenamefont {Harmans},\ and\
		\citenamefont {Mooij}}]{fluxqubitchiorescu2003}%
	\BibitemOpen
	\bibfield  {author} {\bibinfo {author} {\bibfnamefont {I.}~\bibnamefont
			{Chiorescu}}, \bibinfo {author} {\bibfnamefont {Y.}~\bibnamefont {Nakamura}},
		\bibinfo {author} {\bibfnamefont {C.~M.}\ \bibnamefont {Harmans}},\ and\
		\bibinfo {author} {\bibfnamefont {J.}~\bibnamefont {Mooij}},\ }\bibfield
	{title} {\bibinfo {title} {Coherent quantum dynamics of a superconducting
			flux qubit},\ }\href@noop {} {\bibfield  {journal} {\bibinfo  {journal}
			{Science}\ }\textbf {\bibinfo {volume} {299}},\ \bibinfo {pages} {1869}
		(\bibinfo {year} {2003})}\BibitemShut {NoStop}%
	\bibitem [{\citenamefont {Manucharyan}\ \emph {et~al.}(2009)\citenamefont
		{Manucharyan}, \citenamefont {Koch}, \citenamefont {Glazman},\ and\
		\citenamefont {Devoret}}]{fluxoniummanucharyan2009}%
	\BibitemOpen
	\bibfield  {author} {\bibinfo {author} {\bibfnamefont {V.~E.}\ \bibnamefont
			{Manucharyan}}, \bibinfo {author} {\bibfnamefont {J.}~\bibnamefont {Koch}},
		\bibinfo {author} {\bibfnamefont {L.~I.}\ \bibnamefont {Glazman}},\ and\
		\bibinfo {author} {\bibfnamefont {M.~H.}\ \bibnamefont {Devoret}},\
	}\bibfield  {title} {\bibinfo {title} {Fluxonium: Single cooper-pair circuit
			free of charge offsets},\ }\href@noop {} {\bibfield  {journal} {\bibinfo
			{journal} {Science}\ }\textbf {\bibinfo {volume} {326}},\ \bibinfo {pages}
		{113} (\bibinfo {year} {2009})}\BibitemShut {NoStop}%
	\bibitem [{\citenamefont {Koch}\ \emph {et~al.}(2007)\citenamefont {Koch},
		\citenamefont {Terri}, \citenamefont {Gambetta}, \citenamefont {Houck},
		\citenamefont {Schuster}, \citenamefont {Majer}, \citenamefont {Blais},
		\citenamefont {Devoret}, \citenamefont {Girvin},\ and\ \citenamefont
		{Schoelkopf}}]{KochTransmon}%
	\BibitemOpen
	\bibfield  {author} {\bibinfo {author} {\bibfnamefont {J.}~\bibnamefont
			{Koch}}, \bibinfo {author} {\bibfnamefont {M.~Y.}\ \bibnamefont {Terri}},
		\bibinfo {author} {\bibfnamefont {J.}~\bibnamefont {Gambetta}}, \bibinfo
		{author} {\bibfnamefont {A.~A.}\ \bibnamefont {Houck}}, \bibinfo {author}
		{\bibfnamefont {D.}~\bibnamefont {Schuster}}, \bibinfo {author}
		{\bibfnamefont {J.}~\bibnamefont {Majer}}, \bibinfo {author} {\bibfnamefont
			{A.}~\bibnamefont {Blais}}, \bibinfo {author} {\bibfnamefont {M.~H.}\
			\bibnamefont {Devoret}}, \bibinfo {author} {\bibfnamefont {S.~M.}\
			\bibnamefont {Girvin}},\ and\ \bibinfo {author} {\bibfnamefont {R.~J.}\
			\bibnamefont {Schoelkopf}},\ }\bibfield  {title} {\bibinfo {title}
		{Charge-insensitive qubit design derived from the cooper pair box},\
	}\href@noop {} {\bibfield  {journal} {\bibinfo  {journal} {Physical Review
				A}\ }\textbf {\bibinfo {volume} {76}},\ \bibinfo {pages} {042319} (\bibinfo
		{year} {2007})}\BibitemShut {NoStop}%
	\bibitem [{\citenamefont {Chen}\ \emph
		{et~al.}(2014{\natexlab{b}})\citenamefont {Chen}, \citenamefont {Neill},
		\citenamefont {Roushan}, \citenamefont {Leung}, \citenamefont {Fang},
		\citenamefont {Barends}, \citenamefont {Kelly}, \citenamefont {Campbell},
		\citenamefont {Chen}, \citenamefont {Chiaro} \emph {et~al.}}]{chen2014qubit}%
	\BibitemOpen
	\bibfield  {author} {\bibinfo {author} {\bibfnamefont {Y.}~\bibnamefont
			{Chen}}, \bibinfo {author} {\bibfnamefont {C.}~\bibnamefont {Neill}},
		\bibinfo {author} {\bibfnamefont {P.}~\bibnamefont {Roushan}}, \bibinfo
		{author} {\bibfnamefont {N.}~\bibnamefont {Leung}}, \bibinfo {author}
		{\bibfnamefont {M.}~\bibnamefont {Fang}}, \bibinfo {author} {\bibfnamefont
			{R.}~\bibnamefont {Barends}}, \bibinfo {author} {\bibfnamefont
			{J.}~\bibnamefont {Kelly}}, \bibinfo {author} {\bibfnamefont
			{B.}~\bibnamefont {Campbell}}, \bibinfo {author} {\bibfnamefont
			{Z.}~\bibnamefont {Chen}}, \bibinfo {author} {\bibfnamefont {B.}~\bibnamefont
			{Chiaro}}, \emph {et~al.},\ }\bibfield  {title} {\bibinfo {title} {Qubit
			architecture with high coherence and fast tunable coupling},\ }\href@noop {}
	{\bibfield  {journal} {\bibinfo  {journal} {Physical review letters}\
		}\textbf {\bibinfo {volume} {113}},\ \bibinfo {pages} {220502} (\bibinfo
		{year} {2014}{\natexlab{b}})}\BibitemShut {NoStop}%
	\bibitem [{\citenamefont {Zhao}\ \emph {et~al.}(2020)\citenamefont {Zhao},
		\citenamefont {Park}, \citenamefont {Zhao}, \citenamefont {Bal},
		\citenamefont {McRae}, \citenamefont {Long},\ and\ \citenamefont
		{Pappas}}]{mergemonzhao2020}%
	\BibitemOpen
	\bibfield  {author} {\bibinfo {author} {\bibfnamefont {R.}~\bibnamefont
			{Zhao}}, \bibinfo {author} {\bibfnamefont {S.}~\bibnamefont {Park}}, \bibinfo
		{author} {\bibfnamefont {T.}~\bibnamefont {Zhao}}, \bibinfo {author}
		{\bibfnamefont {M.}~\bibnamefont {Bal}}, \bibinfo {author} {\bibfnamefont
			{C.}~\bibnamefont {McRae}}, \bibinfo {author} {\bibfnamefont
			{J.}~\bibnamefont {Long}},\ and\ \bibinfo {author} {\bibfnamefont
			{D.}~\bibnamefont {Pappas}},\ }\bibfield  {title} {\bibinfo {title}
		{Merged-element transmon},\ }\href@noop {} {\bibfield  {journal} {\bibinfo
			{journal} {Physical Review Applied}\ }\textbf {\bibinfo {volume} {14}},\
		\bibinfo {pages} {064006} (\bibinfo {year} {2020})}\BibitemShut {NoStop}%
	\bibitem [{\citenamefont {Mamin}\ \emph {et~al.}(2021)\citenamefont {Mamin},
		\citenamefont {Huang}, \citenamefont {Carnevale}, \citenamefont {Rettner},
		\citenamefont {Arellano}, \citenamefont {Sherwood}, \citenamefont {Kurter},
		\citenamefont {Trimm}, \citenamefont {Sandberg}, \citenamefont {Shelby} \emph
		{et~al.}}]{mergemonmamin2021}%
	\BibitemOpen
	\bibfield  {author} {\bibinfo {author} {\bibfnamefont {H.}~\bibnamefont
			{Mamin}}, \bibinfo {author} {\bibfnamefont {E.}~\bibnamefont {Huang}},
		\bibinfo {author} {\bibfnamefont {S.}~\bibnamefont {Carnevale}}, \bibinfo
		{author} {\bibfnamefont {C.~T.}\ \bibnamefont {Rettner}}, \bibinfo {author}
		{\bibfnamefont {N.}~\bibnamefont {Arellano}}, \bibinfo {author}
		{\bibfnamefont {M.}~\bibnamefont {Sherwood}}, \bibinfo {author}
		{\bibfnamefont {C.}~\bibnamefont {Kurter}}, \bibinfo {author} {\bibfnamefont
			{B.}~\bibnamefont {Trimm}}, \bibinfo {author} {\bibfnamefont
			{M.}~\bibnamefont {Sandberg}}, \bibinfo {author} {\bibfnamefont {R.~M.}\
			\bibnamefont {Shelby}}, \emph {et~al.},\ }\bibfield  {title} {\bibinfo
		{title} {Merged-element transmons: Design and qubit performance},\
	}\href@noop {} {\bibfield  {journal} {\bibinfo  {journal} {Physical Review
				Applied}\ }\textbf {\bibinfo {volume} {16}},\ \bibinfo {pages} {024023}
		(\bibinfo {year} {2021})}\BibitemShut {NoStop}%
	\bibitem [{\citenamefont {Wang}\ \emph {et~al.}(2015)\citenamefont {Wang},
		\citenamefont {Axline}, \citenamefont {Gao}, \citenamefont {Brecht},
		\citenamefont {Chu}, \citenamefont {Frunzio}, \citenamefont {Devoret},\ and\
		\citenamefont {Schoelkopf}}]{Wang2015}%
	\BibitemOpen
	\bibfield  {author} {\bibinfo {author} {\bibfnamefont {C.}~\bibnamefont
			{Wang}}, \bibinfo {author} {\bibfnamefont {C.}~\bibnamefont {Axline}},
		\bibinfo {author} {\bibfnamefont {Y.~Y.}\ \bibnamefont {Gao}}, \bibinfo
		{author} {\bibfnamefont {T.}~\bibnamefont {Brecht}}, \bibinfo {author}
		{\bibfnamefont {Y.}~\bibnamefont {Chu}}, \bibinfo {author} {\bibfnamefont
			{L.}~\bibnamefont {Frunzio}}, \bibinfo {author} {\bibfnamefont
			{M.}~\bibnamefont {Devoret}},\ and\ \bibinfo {author} {\bibfnamefont {R.~J.}\
			\bibnamefont {Schoelkopf}},\ }\bibfield  {title} {\bibinfo {title} {Surface
			participation and dielectric loss in superconducting qubits},\ }\href@noop {}
	{\bibfield  {journal} {\bibinfo  {journal} {Applied Physics Letters}\
		}\textbf {\bibinfo {volume} {107}},\ \bibinfo {pages} {162601} (\bibinfo
		{year} {2015})}\BibitemShut {NoStop}%
	\bibitem [{\citenamefont {Lisenfeld}\ \emph {et~al.}(2019)\citenamefont
		{Lisenfeld}, \citenamefont {Bilmes}, \citenamefont {Megrant}, \citenamefont
		{Barends}, \citenamefont {Kelly}, \citenamefont {Klimov}, \citenamefont
		{Weiss}, \citenamefont {Martinis},\ and\ \citenamefont
		{Ustinov}}]{Lisenfeld19}%
	\BibitemOpen
	\bibfield  {author} {\bibinfo {author} {\bibfnamefont {J.}~\bibnamefont
			{Lisenfeld}}, \bibinfo {author} {\bibfnamefont {A.}~\bibnamefont {Bilmes}},
		\bibinfo {author} {\bibfnamefont {A.}~\bibnamefont {Megrant}}, \bibinfo
		{author} {\bibfnamefont {R.}~\bibnamefont {Barends}}, \bibinfo {author}
		{\bibfnamefont {J.}~\bibnamefont {Kelly}}, \bibinfo {author} {\bibfnamefont
			{P.}~\bibnamefont {Klimov}}, \bibinfo {author} {\bibfnamefont
			{G.}~\bibnamefont {Weiss}}, \bibinfo {author} {\bibfnamefont {J.~M.}\
			\bibnamefont {Martinis}},\ and\ \bibinfo {author} {\bibfnamefont {A.~V.}\
			\bibnamefont {Ustinov}},\ }\bibfield  {title} {\bibinfo {title} {Electric
			field spectroscopy of material defects in transmon qubits},\ }\href@noop {}
	{\bibfield  {journal} {\bibinfo  {journal} {npj Quantum Information}\
		}\textbf {\bibinfo {volume} {5}},\ \bibinfo {pages} {1} (\bibinfo {year}
		{2019})}\BibitemShut {NoStop}%
	\bibitem [{\citenamefont {Bilmes}\ \emph {et~al.}(2020)\citenamefont {Bilmes},
		\citenamefont {Megrant}, \citenamefont {Klimov}, \citenamefont {Weiss},
		\citenamefont {Martinis}, \citenamefont {Ustinov},\ and\ \citenamefont
		{Lisenfeld}}]{Bilmes20}%
	\BibitemOpen
	\bibfield  {author} {\bibinfo {author} {\bibfnamefont {A.}~\bibnamefont
			{Bilmes}}, \bibinfo {author} {\bibfnamefont {A.}~\bibnamefont {Megrant}},
		\bibinfo {author} {\bibfnamefont {P.}~\bibnamefont {Klimov}}, \bibinfo
		{author} {\bibfnamefont {G.}~\bibnamefont {Weiss}}, \bibinfo {author}
		{\bibfnamefont {J.~M.}\ \bibnamefont {Martinis}}, \bibinfo {author}
		{\bibfnamefont {A.~V.}\ \bibnamefont {Ustinov}},\ and\ \bibinfo {author}
		{\bibfnamefont {J.}~\bibnamefont {Lisenfeld}},\ }\bibfield  {title} {\bibinfo
		{title} {Resolving the positions of defects in superconducting quantum
			bits},\ }\href@noop {} {\bibfield  {journal} {\bibinfo  {journal} {Scientific
				Reports}\ }\textbf {\bibinfo {volume} {10}},\ \bibinfo {pages} {1} (\bibinfo
		{year} {2020})}\BibitemShut {NoStop}%
	\bibitem [{\citenamefont {Murray}(2021)}]{murray2021}%
	\BibitemOpen
	\bibfield  {author} {\bibinfo {author} {\bibfnamefont {C.~E.}\ \bibnamefont
			{Murray}},\ }\bibfield  {title} {\bibinfo {title} {Material matters in
			superconducting qubits},\ }\href@noop {} {\bibfield  {journal} {\bibinfo
			{journal} {Materials Science and Engineering: R: Reports}\ }\textbf {\bibinfo
			{volume} {146}},\ \bibinfo {pages} {100646} (\bibinfo {year}
		{2021})}\BibitemShut {NoStop}%
	\bibitem [{\citenamefont {Van~Harlingen}\ \emph {et~al.}(2004)\citenamefont
		{Van~Harlingen}, \citenamefont {Robertson}, \citenamefont {Plourde},
		\citenamefont {Reichardt}, \citenamefont {Crane},\ and\ \citenamefont
		{Clarke}}]{HarlingenSQUID}%
	\BibitemOpen
	\bibfield  {author} {\bibinfo {author} {\bibfnamefont {D.~J.}\ \bibnamefont
			{Van~Harlingen}}, \bibinfo {author} {\bibfnamefont {T.~L.}\ \bibnamefont
			{Robertson}}, \bibinfo {author} {\bibfnamefont {B.~L.~T.}\ \bibnamefont
			{Plourde}}, \bibinfo {author} {\bibfnamefont {P.~A.}\ \bibnamefont
			{Reichardt}}, \bibinfo {author} {\bibfnamefont {T.~A.}\ \bibnamefont
			{Crane}},\ and\ \bibinfo {author} {\bibfnamefont {J.}~\bibnamefont
			{Clarke}},\ }\bibfield  {title} {\bibinfo {title} {Decoherence in
			josephson-junction qubits due to critical-current fluctuations},\ }\href
	{https://doi.org/10.1103/PhysRevB.70.064517} {\bibfield  {journal} {\bibinfo
			{journal} {Phys. Rev. B}\ }\textbf {\bibinfo {volume} {70}},\ \bibinfo
		{pages} {064517} (\bibinfo {year} {2004})}\BibitemShut {NoStop}%
	\bibitem [{\citenamefont {Likharev}(1981)}]{LIKHAREV19811079}%
	\BibitemOpen
	\bibfield  {author} {\bibinfo {author} {\bibfnamefont {K.}~\bibnamefont
			{Likharev}},\ }\bibfield  {title} {\bibinfo {title} {Macroscopic quantum
			tunneling and dissipation in josephson junctions and squids},\ }\href
	{https://doi.org/https://doi.org/10.1016/0378-4363(81)90843-3} {\bibfield
		{journal} {\bibinfo  {journal} {Physica B+C}\ }\textbf {\bibinfo {volume}
			{108}},\ \bibinfo {pages} {1079} (\bibinfo {year} {1981})}\BibitemShut
	{NoStop}%
	\bibitem [{\citenamefont {Lang}\ \emph {et~al.}(2003)\citenamefont {Lang},
		\citenamefont {Nam}, \citenamefont {Aumentado}, \citenamefont {Urbina},\ and\
		\citenamefont {Martinis}}]{phasequbitmartinis2003}%
	\BibitemOpen
	\bibfield  {author} {\bibinfo {author} {\bibfnamefont {K.}~\bibnamefont
			{Lang}}, \bibinfo {author} {\bibfnamefont {S.}~\bibnamefont {Nam}}, \bibinfo
		{author} {\bibfnamefont {J.}~\bibnamefont {Aumentado}}, \bibinfo {author}
		{\bibfnamefont {C.}~\bibnamefont {Urbina}},\ and\ \bibinfo {author}
		{\bibfnamefont {J.~M.}\ \bibnamefont {Martinis}},\ }\bibfield  {title}
	{\bibinfo {title} {Banishing quasiparticles from josephson-junction qubits:
			Why and how to do it},\ }\href@noop {} {\bibfield  {journal} {\bibinfo
			{journal} {IEEE Transactions on Applied Superconductivity}\ }\textbf
		{\bibinfo {volume} {13}},\ \bibinfo {pages} {989} (\bibinfo {year}
		{2003})}\BibitemShut {NoStop}%
	\bibitem [{\citenamefont {Martinis}(2009)}]{phasequbitmartinis2009}%
	\BibitemOpen
	\bibfield  {author} {\bibinfo {author} {\bibfnamefont {J.~M.}\ \bibnamefont
			{Martinis}},\ }\bibfield  {title} {\bibinfo {title} {Superconducting phase
			qubits},\ }\href@noop {} {\bibfield  {journal} {\bibinfo  {journal} {Quantum
				information processing}\ }\textbf {\bibinfo {volume} {8}},\ \bibinfo {pages}
		{81} (\bibinfo {year} {2009})}\BibitemShut {NoStop}%
	\bibitem [{\citenamefont {Lisenfeld}(2008)}]{phasequbitlisenfeld2008}%
	\BibitemOpen
	\bibfield  {author} {\bibinfo {author} {\bibfnamefont {J.}~\bibnamefont
			{Lisenfeld}},\ }\href@noop {} {\emph {\bibinfo {title} {Experiments on
				superconducting Josephson phase quantum bits}}}\ (\bibinfo  {publisher}
	{Citeseer},\ \bibinfo {year} {2008})\BibitemShut {NoStop}%
	\bibitem [{\citenamefont {Castellano}\ \emph {et~al.}(2006)\citenamefont
		{Castellano}, \citenamefont {Chiarello}, \citenamefont {Torrioli},\ and\
		\citenamefont {Carelli}}]{RFSQUIDcastellano2006}%
	\BibitemOpen
	\bibfield  {author} {\bibinfo {author} {\bibfnamefont {M.~G.}\ \bibnamefont
			{Castellano}}, \bibinfo {author} {\bibfnamefont {F.}~\bibnamefont
			{Chiarello}}, \bibinfo {author} {\bibfnamefont {G.}~\bibnamefont
			{Torrioli}},\ and\ \bibinfo {author} {\bibfnamefont {P.}~\bibnamefont
			{Carelli}},\ }\bibfield  {title} {\bibinfo {title} {Static flux bias of a
			flux qubit using persistent current trapping},\ }\href@noop {} {\bibfield
		{journal} {\bibinfo  {journal} {Superconductor Science and Technology}\
		}\textbf {\bibinfo {volume} {19}},\ \bibinfo {pages} {1158} (\bibinfo {year}
		{2006})}\BibitemShut {NoStop}%
	\bibitem [{\citenamefont {Jiang}\ \emph {et~al.}(2024)\citenamefont {Jiang},
		\citenamefont {Xu}, \citenamefont {Li}, \citenamefont {Yan}, \citenamefont
		{Gong}, \citenamefont {Rong}, \citenamefont {Sun}, \citenamefont {Sun},
		\citenamefont {Jiang}, \citenamefont {Deng} \emph {et~al.}}]{rfSQUIDtuning}%
	\BibitemOpen
	\bibfield  {author} {\bibinfo {author} {\bibfnamefont {L.}~\bibnamefont
			{Jiang}}, \bibinfo {author} {\bibfnamefont {Y.}~\bibnamefont {Xu}}, \bibinfo
		{author} {\bibfnamefont {S.}~\bibnamefont {Li}}, \bibinfo {author}
		{\bibfnamefont {Z.}~\bibnamefont {Yan}}, \bibinfo {author} {\bibfnamefont
			{M.}~\bibnamefont {Gong}}, \bibinfo {author} {\bibfnamefont {T.}~\bibnamefont
			{Rong}}, \bibinfo {author} {\bibfnamefont {C.}~\bibnamefont {Sun}}, \bibinfo
		{author} {\bibfnamefont {T.}~\bibnamefont {Sun}}, \bibinfo {author}
		{\bibfnamefont {T.}~\bibnamefont {Jiang}}, \bibinfo {author} {\bibfnamefont
			{H.}~\bibnamefont {Deng}}, \emph {et~al.},\ }\bibfield  {title} {\bibinfo
		{title} {In situ qubit frequency tuning circuit for scalable superconducting
			quantum computing: Scheme and experiment},\ }\href@noop {} {\bibfield
		{journal} {\bibinfo  {journal} {arXiv preprint arXiv:2407.21415}\ } (\bibinfo
		{year} {2024})}\BibitemShut {NoStop}%
	\bibitem [{\citenamefont {Marchiori}\ \emph {et~al.}(2022)\citenamefont
		{Marchiori}, \citenamefont {Ceccarelli}, \citenamefont {Rossi}, \citenamefont
		{Romagnoli}, \citenamefont {Herrmann}, \citenamefont {Besse}, \citenamefont
		{Krinner}, \citenamefont {Wallraff},\ and\ \citenamefont
		{Poggio}}]{VortexHopping}%
	\BibitemOpen
	\bibfield  {author} {\bibinfo {author} {\bibfnamefont {E.}~\bibnamefont
			{Marchiori}}, \bibinfo {author} {\bibfnamefont {L.}~\bibnamefont
			{Ceccarelli}}, \bibinfo {author} {\bibfnamefont {N.}~\bibnamefont {Rossi}},
		\bibinfo {author} {\bibfnamefont {G.}~\bibnamefont {Romagnoli}}, \bibinfo
		{author} {\bibfnamefont {J.}~\bibnamefont {Herrmann}}, \bibinfo {author}
		{\bibfnamefont {J.-C.}\ \bibnamefont {Besse}}, \bibinfo {author}
		{\bibfnamefont {S.}~\bibnamefont {Krinner}}, \bibinfo {author} {\bibfnamefont
			{A.}~\bibnamefont {Wallraff}},\ and\ \bibinfo {author} {\bibfnamefont
			{M.}~\bibnamefont {Poggio}},\ }\bibfield  {title} {\bibinfo {title} {Magnetic
			imaging of superconducting qubit devices with scanning squid-on-tip},\
	}\href@noop {} {\bibfield  {journal} {\bibinfo  {journal} {Applied Physics
				Letters}\ }\textbf {\bibinfo {volume} {121}} (\bibinfo {year}
		{2022})}\BibitemShut {NoStop}%
	\bibitem [{\citenamefont {Dunsworth}\ \emph {et~al.}(2017)\citenamefont
		{Dunsworth}, \citenamefont {Megrant}, \citenamefont {Quintana}, \citenamefont
		{Chen}, \citenamefont {Barends}, \citenamefont {Burkett}, \citenamefont
		{Foxen}, \citenamefont {Chen}, \citenamefont {Chiaro}, \citenamefont {Fowler}
		\emph {et~al.}}]{dunsworth2017}%
	\BibitemOpen
	\bibfield  {author} {\bibinfo {author} {\bibfnamefont {A.}~\bibnamefont
			{Dunsworth}}, \bibinfo {author} {\bibfnamefont {A.}~\bibnamefont {Megrant}},
		\bibinfo {author} {\bibfnamefont {C.}~\bibnamefont {Quintana}}, \bibinfo
		{author} {\bibfnamefont {Z.}~\bibnamefont {Chen}}, \bibinfo {author}
		{\bibfnamefont {R.}~\bibnamefont {Barends}}, \bibinfo {author} {\bibfnamefont
			{B.}~\bibnamefont {Burkett}}, \bibinfo {author} {\bibfnamefont
			{B.}~\bibnamefont {Foxen}}, \bibinfo {author} {\bibfnamefont
			{Y.}~\bibnamefont {Chen}}, \bibinfo {author} {\bibfnamefont {B.}~\bibnamefont
			{Chiaro}}, \bibinfo {author} {\bibfnamefont {A.}~\bibnamefont {Fowler}},
		\emph {et~al.},\ }\bibfield  {title} {\bibinfo {title} {Characterization and
			reduction of capacitive loss induced by sub-micron josephson junction
			fabrication in superconducting qubits},\ }\href@noop {} {\bibfield  {journal}
		{\bibinfo  {journal} {Applied Physics Letters}\ }\textbf {\bibinfo {volume}
			{111}},\ \bibinfo {pages} {022601} (\bibinfo {year} {2017})}\BibitemShut
	{NoStop}%
	\bibitem [{\citenamefont {Rosenberg}\ \emph {et~al.}(2017)\citenamefont
		{Rosenberg}, \citenamefont {Kim}, \citenamefont {Das}, \citenamefont {Yost},
		\citenamefont {Gustavsson}, \citenamefont {Hover}, \citenamefont {Krantz},
		\citenamefont {Melville}, \citenamefont {Racz}, \citenamefont {Samach} \emph
		{et~al.}}]{flipchiprosenberg20173}%
	\BibitemOpen
	\bibfield  {author} {\bibinfo {author} {\bibfnamefont {D.}~\bibnamefont
			{Rosenberg}}, \bibinfo {author} {\bibfnamefont {D.}~\bibnamefont {Kim}},
		\bibinfo {author} {\bibfnamefont {R.}~\bibnamefont {Das}}, \bibinfo {author}
		{\bibfnamefont {D.}~\bibnamefont {Yost}}, \bibinfo {author} {\bibfnamefont
			{S.}~\bibnamefont {Gustavsson}}, \bibinfo {author} {\bibfnamefont
			{D.}~\bibnamefont {Hover}}, \bibinfo {author} {\bibfnamefont
			{P.}~\bibnamefont {Krantz}}, \bibinfo {author} {\bibfnamefont
			{A.}~\bibnamefont {Melville}}, \bibinfo {author} {\bibfnamefont
			{L.}~\bibnamefont {Racz}}, \bibinfo {author} {\bibfnamefont {G.}~\bibnamefont
			{Samach}}, \emph {et~al.},\ }\bibfield  {title} {\bibinfo {title} {3d
			integrated superconducting qubits},\ }\href@noop {} {\bibfield  {journal}
		{\bibinfo  {journal} {npj quantum information}\ }\textbf {\bibinfo {volume}
			{3}},\ \bibinfo {pages} {42} (\bibinfo {year} {2017})}\BibitemShut {NoStop}%
	\bibitem [{\citenamefont {Braginski}(2019)}]{reviewbraginski2019}%
	\BibitemOpen
	\bibfield  {author} {\bibinfo {author} {\bibfnamefont {A.~I.}\ \bibnamefont
			{Braginski}},\ }\bibfield  {title} {\bibinfo {title} {Superconductor
			electronics: Status and outlook},\ }\href@noop {} {\bibfield  {journal}
		{\bibinfo  {journal} {Journal of superconductivity and novel magnetism}\
		}\textbf {\bibinfo {volume} {32}},\ \bibinfo {pages} {23} (\bibinfo {year}
		{2019})}\BibitemShut {NoStop}%
	\bibitem [{\citenamefont {Macklin}\ \emph {et~al.}(2015)\citenamefont
		{Macklin}, \citenamefont {O’brien}, \citenamefont {Hover}, \citenamefont
		{Schwartz}, \citenamefont {Bolkhovsky}, \citenamefont {Zhang}, \citenamefont
		{Oliver},\ and\ \citenamefont {Siddiqi}}]{twpamacklin2015near}%
	\BibitemOpen
	\bibfield  {author} {\bibinfo {author} {\bibfnamefont {C.}~\bibnamefont
			{Macklin}}, \bibinfo {author} {\bibfnamefont {K.}~\bibnamefont {O’brien}},
		\bibinfo {author} {\bibfnamefont {D.}~\bibnamefont {Hover}}, \bibinfo
		{author} {\bibfnamefont {M.}~\bibnamefont {Schwartz}}, \bibinfo {author}
		{\bibfnamefont {V.}~\bibnamefont {Bolkhovsky}}, \bibinfo {author}
		{\bibfnamefont {X.}~\bibnamefont {Zhang}}, \bibinfo {author} {\bibfnamefont
			{W.}~\bibnamefont {Oliver}},\ and\ \bibinfo {author} {\bibfnamefont
			{I.}~\bibnamefont {Siddiqi}},\ }\bibfield  {title} {\bibinfo {title} {A
			near--quantum-limited josephson traveling-wave parametric amplifier},\
	}\href@noop {} {\bibfield  {journal} {\bibinfo  {journal} {Science}\ }\textbf
		{\bibinfo {volume} {350}},\ \bibinfo {pages} {307} (\bibinfo {year}
		{2015})}\BibitemShut {NoStop}%
	\bibitem [{\citenamefont {Aumentado}(2020)}]{twpareviewaumentado2020}%
	\BibitemOpen
	\bibfield  {author} {\bibinfo {author} {\bibfnamefont {J.}~\bibnamefont
			{Aumentado}},\ }\bibfield  {title} {\bibinfo {title} {Superconducting
			parametric amplifiers: The state of the art in josephson parametric
			amplifiers},\ }\href@noop {} {\bibfield  {journal} {\bibinfo  {journal} {IEEE
				Microwave magazine}\ }\textbf {\bibinfo {volume} {21}},\ \bibinfo {pages}
		{45} (\bibinfo {year} {2020})}\BibitemShut {NoStop}%
	\bibitem [{\citenamefont {Bunyk}\ \emph {et~al.}(2001)\citenamefont {Bunyk},
		\citenamefont {Likharev},\ and\ \citenamefont {Zinoviev}}]{rsfqbunykreview}%
	\BibitemOpen
	\bibfield  {author} {\bibinfo {author} {\bibfnamefont {P.}~\bibnamefont
			{Bunyk}}, \bibinfo {author} {\bibfnamefont {K.}~\bibnamefont {Likharev}},\
		and\ \bibinfo {author} {\bibfnamefont {D.}~\bibnamefont {Zinoviev}},\
	}\bibfield  {title} {\bibinfo {title} {Rsfq technology: Physics and
			devices},\ }\href@noop {} {\bibfield  {journal} {\bibinfo  {journal}
			{International journal of high speed electronics and systems}\ }\textbf
		{\bibinfo {volume} {11}},\ \bibinfo {pages} {257} (\bibinfo {year}
		{2001})}\BibitemShut {NoStop}%
	\bibitem [{\citenamefont {Mazin}(2009)}]{MKIDreviewmazin2009}%
	\BibitemOpen
	\bibfield  {author} {\bibinfo {author} {\bibfnamefont {B.~A.}\ \bibnamefont
			{Mazin}},\ }\bibfield  {title} {\bibinfo {title} {Microwave kinetic
			inductance detectors: The first decade},\ }in\ \href@noop {} {\emph {\bibinfo
			{booktitle} {AIP Conference Proceedings}}},\ Vol.\ \bibinfo {volume} {1185}\
	(\bibinfo {organization} {American Institute of Physics},\ \bibinfo {year}
	{2009})\ pp.\ \bibinfo {pages} {135--142}\BibitemShut {NoStop}%
	\bibitem [{\citenamefont {Oates}(2022)}]{filtersoates2022}%
	\BibitemOpen
	\bibfield  {author} {\bibinfo {author} {\bibfnamefont {D.~E.}\ \bibnamefont
			{Oates}},\ }\bibfield  {title} {\bibinfo {title} {Microwave resonators and
			filters},\ }in\ \href@noop {} {\emph {\bibinfo {booktitle} {Handbook of
				Superconductivity}}}\ (\bibinfo  {publisher} {CRC Press},\ \bibinfo {year}
	{2022})\ pp.\ \bibinfo {pages} {596--608}\BibitemShut {NoStop}%
	\bibitem [{\citenamefont {Rosa}(1908)}]{rosa1908self}%
	\BibitemOpen
	\bibfield  {author} {\bibinfo {author} {\bibfnamefont {E.~B.}\ \bibnamefont
			{Rosa}},\ }\bibfield  {title} {\bibinfo {title} {The self and mutual
			inductances of linear conductors},\ }\href
	{https://nvlpubs.nist.gov/nistpubs/bulletin/04/nbsbulletinv4n2p301_A2b.pdf}
	{\bibfield  {journal} {\bibinfo  {journal} {Bulletin of the Bureau of
				Standards}\ }\textbf {\bibinfo {volume} {4}},\ \bibinfo {pages} {301}
		(\bibinfo {year} {1908})},\ \bibinfo {note} {scientific Paper No.
		80}\BibitemShut {NoStop}%
	\bibitem [{Pra(2004)}]{Prasad2004}%
	\BibitemOpen
	\bibinfo {title} {Process technology},\ in\ \href
	{https://doi.org/10.1007/1-4020-7763-7_4} {\emph {\bibinfo {booktitle}
			{Advanced Wirebond Interconnection Technology}}}\ (\bibinfo  {publisher}
	{Springer US},\ \bibinfo {address} {Boston, MA},\ \bibinfo {year} {2004})\
	pp.\ \bibinfo {pages} {163--480}\BibitemShut {NoStop}%
	\bibitem [{\citenamefont {Bilmes}\ \emph {et~al.}(2021)\citenamefont {Bilmes},
		\citenamefont {H{\"a}ndel}, \citenamefont {Volosheniuk}, \citenamefont
		{Ustinov},\ and\ \citenamefont {Lisenfeld}}]{Bilmes21}%
	\BibitemOpen
	\bibfield  {author} {\bibinfo {author} {\bibfnamefont {A.}~\bibnamefont
			{Bilmes}}, \bibinfo {author} {\bibfnamefont {A.~K.}\ \bibnamefont
			{H{\"a}ndel}}, \bibinfo {author} {\bibfnamefont {S.}~\bibnamefont
			{Volosheniuk}}, \bibinfo {author} {\bibfnamefont {A.~V.}\ \bibnamefont
			{Ustinov}},\ and\ \bibinfo {author} {\bibfnamefont {J.}~\bibnamefont
			{Lisenfeld}},\ }\bibfield  {title} {\bibinfo {title} {In-situ bandaged
			josephson junctions for superconducting quantum processors},\ }\href@noop {}
	{\bibfield  {journal} {\bibinfo  {journal} {Superconductor Science and
				Technology}\ }\textbf {\bibinfo {volume} {34}},\ \bibinfo {pages} {125011}
		(\bibinfo {year} {2021})}\BibitemShut {NoStop}%
	\bibitem [{\citenamefont {Daum}\ \emph {et~al.}(2025)\citenamefont {Daum},
		\citenamefont {Berlitz}, \citenamefont {Ustinov},\ and\ \citenamefont
		{Lisenfeld}}]{Daum25}%
	\BibitemOpen
	\bibfield  {author} {\bibinfo {author} {\bibfnamefont {E.}~\bibnamefont
			{Daum}}, \bibinfo {author} {\bibfnamefont {B.}~\bibnamefont {Berlitz}},
		\bibinfo {author} {\bibfnamefont {A.}~\bibnamefont {Ustinov}},\ and\ \bibinfo
		{author} {\bibfnamefont {J.}~\bibnamefont {Lisenfeld}},\ }\bibfield  {title}
	{\bibinfo {title} {Two-level-systems in coherent merged-element-qubits}}
	(\bibinfo {year} {2025}),\ \bibinfo {note} {in preparation.}\BibitemShut
	{Stop}%
\end{thebibliography}
\end{document}